\documentstyle[multicol,eqsecnum,aps,graphicx]{revtex}

\draft

\begin{document}
\title{Can the SO(10) Model with Two Higgs Doublets  \\[.2in]
Reproduce the Observed Fermion Masses?}

\author{\bf K.~Matsuda, Y.~Koide$^{(a)}$, and T.~Fukuyama}%
\address{%
Department of Physics, Ritsumeikan University, 
Kusatsu, Shiga, 525-8577 Japan \\
(a) Department of Physics, University of Shizuoka, Shizuoka 422-8526 
Japan}%
\date{\today}

\maketitle
\begin{abstract}
It is usually considered that the SO(10) model with one {\bf 10} and 
one {\bf 126} Higgs scalars cannot reproduce the observed quark and
charged lepton masses.  Against this conventional conjecture, we find 
solutions of the parameters which can give the observed fermion mass 
spectra. 
The SO(10) model with one {\bf 10} and one {\bf 120} Higgs 
scalars is also discussed.
\end{abstract}

\begin{multicols}{2}
\narrowtext
\section{Introduction}

The grand unification theory (GUT) is very attractive as a unified 
description of the fundamental forces in the nature.  
Especially, the SO(10) model is the most attractive to us when we take
the unification of the quarks and leptons into consideration.
However, in order to reproduce the observed quark and lepton masses 
and mixings, usually, a lot of Higgs scalars 
are brought into the model.  We think that the nature is simple.  
What is of the greatest interest to us is to know the minimum number 
of the Higgs scalars which can give the observed fermion mass spectra.  
A model with one Higgs scalar is obviously ruled out for the 
description of the realistic quark and lepton mass spectra.  
Then, how is a model with two different types of Higgs scalars (e.g.,  
{\bf 10} and {\bf 126} scalars)?

In the SO(10) GUT scenario, a model with one {\bf 10} and one {\bf 126} 
Higgs scalars leads to the relation \cite{mohapatra}
\begin{equation}
M_{e}=c_{u}M_{u}+c_{d}M_{d},
\label{eq82514}
\end{equation}
where $M_e$, $M_u$ and $M_d$ are charged lepton, up-quark and down-quark mass 
matrices, respectively.  It is widely accepted that there will be almost no solution 
of $c_u$ and $c_d$ which give the observed fermion mass spectra.  The reason 
is as follows: We take a basis on which the up-quark mass matrix $M_u$ is 
diagonal ($M_u=D_u$).  Then, the relation (\ref{eq82514}) is expressed as
\begin{equation}
\widetilde{M_e}=c_{u}D_{u}+c_{d}\widetilde{M_d}.
\end{equation}
Considering that $\widetilde{M_d}$ is almost diagonal and the mass hierarchy 
of up-quark sector is much severe than that of down-quark sector, we observe 
that the contribution to the first and the second generation part of 
$\widetilde{M_e}$ from the up-quark part $D_u$ is negligible so that it is 
proportional to that of $\widetilde{M_d}$.  
Thus, the relation (\ref{eq82514}) which 
predicts $m_e/m_{\mu}{\simeq}m_d/m_s$ does not reproduce the observed 
hierarchical structure of the down-quark and charged lepton masses \cite{takasugi} such as 
predicted by Georgi-Jarlskog mass relations $m_b=m_{\tau}$, $m_s=m_{\mu}/3$ 
and $m_d=3m_e$ at the GUT scale \cite{georgi}. 
However, the above conclusion is somewhat impatient one.  (i) It is too 
simplified to regard $\widetilde{M_d}$ as almost diagonal.  (ii) We must 
check a possibility that the mass relations are satisfied with the opposite 
signs, i.e., $m_b={\pm}m_{\tau}$, $m_s={\pm}m_{\mu}/3$ and $m_d={\pm}3m_e$.  
(iii) The mass values at the GUT scale which are evaluated from the observed 
values by using the renormalization group equations show sizable deviations 
from the Georgi-Jarskog relations.  The purpose of the present paper is to 
investigate systematically whether there are solutions of $c_u$ and $c_d$ 
which give the realistic quark and lepton masses or not.  

\section{Outline of the investigation}

In the SO(10) GUT model with one {\bf 10} and one {\bf 126} Higgs scalars, 
the down-quark and down-lepton mass matrices $M_d$ and $M_e$ are given by 
\begin{equation}
M_d=M_0+M_1,\ \ M_e=M_0- 3 M_1, \label{eq90101}
\end{equation}
where $M_0$ and $M_1$ are mass matrices which are generated by the {\bf 10} 
and {\bf 126} Higgs scalars $\phi_{10}$ and $\phi_{126}$, respectively.  
Inversely, we obtain
\begin{equation}
M_0={\frac{1}{4}}(3M_d+M_e),\ \ M_1={\frac{1}{4}}(M_d-M_e). \label{eq82513}
\end{equation}
On the other hand, the up-quark mass matrix $M_u$ is given by 
\begin{equation}
M_u=c_0M_0+c_1M_1, 
\end{equation}
where 
\begin{eqnarray}
c_0&=&v_0^u/v_0^d=\langle\phi^{u0}_{10}\rangle/\langle\phi^{d0}_{10}\rangle,
\nonumber \\
c_1&=&v_1^u/v_1^d=\langle\phi^{u0}_{126}\rangle/\langle\phi^{d0}_{126}\rangle, 
\end{eqnarray}
and $\phi^u$ and $\phi^d$ denote Higgs scalar components which couple with 
up- and down-quark sectors, respectively.  Therefore, by using the relations 
Eq.(\ref{eq82513}), we obtain the relation
\begin{equation}
M_e=c_dM_d+c_uM_u, \label{eq82515}
\end{equation}
where
\begin{equation}
c_d=-
{\frac{3c_0+c_1}{c_0-c_1}} \ ,\ \ 
c_u={\frac{4}{c_0-c_1}} \ .
\end{equation}

For convenience, first, we investigate 
the case that the matrices $M_u$, $M_d$ 
and $M_e$ are symmetrical matrices at the unification scale 
because we assume that they are generated 
by the {\bf 10} and {\bf 126} Higgs. 
Then, we can diagonalize 
those by unitary matrices $U_u$, $U_d$ and $U_e$, respectively, as 
\begin{equation}
U_u^{T}M_uU_u=D_u \ , \ \
U_d^{T}M_dU_d=D_d \ , \
U_e^{T}M_eU_e=D_e \ , \label{eq90102}
\end{equation}
where $D_u$, $D_d$ and $D_e$ are diagonal matrices.  Since the 
Cabibbo-Kobayashi-Maskawa (CKM) matrix $V$ is given by 
\begin{equation}
V=U_u^{T} U_d^* \ ,
\end{equation}
the relation (\ref{eq82515}) is re-written as follows: 
\begin{equation}
(U_e^{\dagger}U_u)^T D_e(U_e^{\dagger}U_u)
=c_{d}V D_{d}V^{T}
+c_uD_u .
\end{equation}
At present, we have almost known the experimental values of $D_e$, $D_u$ and 
$V D_d V^{\dagger}$. 
Therefore, we obtain the independent three equations:
\end{multicols}
\widetext
\hspace{-0.5cm}
\rule{8.7cm}{0.1mm}\rule{0.1mm}{2mm}
\begin{eqnarray}
{\rm Tr}D_e D_e^\dagger &=& |c_d|^2 \, {\rm Tr}
\Bigl[(V D_{d}V^{T}+\kappa D_u)
(V D_{d}V^{T}+\kappa D_u)^\dagger\Bigr],
 \label{eq82501}\\
{\rm Tr}(D_e D_e^\dagger)^2 &=&|c_d|^4 \,
{\rm Tr}\Bigl[
((V D_{d}V^{T}+\kappa D_u)
(V D_{d}V^{T}+\kappa D_u)^\dagger)^2\Bigr],
 \label{eq82502}\\
{\rm det}D_eD_e^\dagger &=& |c_d|^6 \, {\rm det}
\Bigl[(V D_{d}V^{T}+\kappa D_u)
(V D_{d}V^{T}+\kappa D_u)^\dagger\Bigr],
\label{eq82503}
\end{eqnarray}
\hspace{9.2cm}
\rule[-2mm]{0.1mm}{2mm}\rule{8.7cm}{0.1mm}
\begin{multicols}{2}
\narrowtext
\hspace{-0.33cm}where $\kappa=c_u/c_d$.  
By eliminating the parameter \(c_d\), we have two equations 
for the parameter \(\kappa\):
\begin{eqnarray}
\frac{(m_e^2+m_\mu^2+m_\tau^2)^3}{m_e^2 m_\mu^2 m_\tau^2}&=&
  \frac{(\ref{eq82501})^3}{(\ref{eq82503})}, \label{eq82511}\\
\frac{(m_e^2+m_\mu^2+m_\tau^2)^2}
     {2(m_e^2 m_\mu^2+m_\mu^2 m_\tau^2+m_\tau^2 m_e^2)}&=&
\frac{(\ref{eq82501})^2}{(\ref{eq82501})^2-(\ref{eq82502})}, \label{eq82512}
\end{eqnarray}
where \((\ref{eq82501})^3\), for instance, 
means the right-hand side of (\ref{eq82501}) to the third power.
Let us denote the parameter values of $\kappa$ evaluated from (\ref{eq82511}) 
and (\ref{eq82512}) as ${\kappa}_A$ and ${\kappa}_B$, respectively.  
If ${\kappa}_A$ and ${\kappa}_B$ coincide with each other, 
then we have a possibility 
that the SO(10) GUT model can reproduce the observed quark and lepton mass 
spectra.  If ${\kappa}_A$ and ${\kappa}_B$ do not so, the SO(10) model with 
one {\bf 10} and one {\bf 126} Higgs scalars is ruled out, and we must bring 
more Higgs scalars into the model.  Of course, in the numerical evaluation, 
the values ${\kappa}_A$ and ${\kappa}_B$ will have sizable errors, because the 
observed values $D_e$, $D_u$, $D_d$ and $V$ have experimental errors, and the 
values at the GUT scale also have errors.  
The values $\kappa_A$ and $\kappa_B$ are not so sensitive to the 
renormalization group equation effect (evolution effect), because those are 
almost determined only by the mass ratios. 
(More details will be discussed in the Sec.~III.)
Therefore, we will evaluate 
${\kappa}_A$ and ${\kappa}_B$ by using the center values at $\mu=m_Z$
in the Sec.~IV.  
If we find $\kappa_A\simeq\kappa_B$, we will give further 
detailed numerical study only for the case.  

\section{Evolution effect}

The relations (2.13) and (2.14) hold only at the unification
scale $\mu=\Lambda_X$ On the other hand, we know only the 
experimental values of the fermion masses $m_f$ and 
CKM matrix parameters $V_{ij}$ at the electroweak 
scale $\mu=m_Z$. 
For a model which does not have any intermediate 
energy scales, we can straightforwardly estimate the values of 
$m_f$ and $V_{ij}$ 
at $\mu=\Lambda_X$ from those at $\mu=m_Z$ 
by the one-loop renormalization equation
\begin{equation}
\frac{dY_f}{dt}=\frac{1}{16\pi^2}(T_f-G_f+H_f)Y_f
\end{equation}
where $T_f$, $G_f$ and $H_f$ denote contributions from fermion-loop 
corrections, vertex corrections due to the gauge bosons and 
vertex corrections due to the Higgs boson(s), respectively.
Therefore, we can directly check the relations (2.13) and (2.14) 
by substituting the observable quantities $m_f$ and $V_{ij}$ 
at $\mu=\Lambda_X$.
However, for a model which has an intermediate energy scale such 
as a non-SUSY model, the values of $m_f$ and $V_{ij}$ at $\mu=
\Lambda_X$ are highly model-dependent, so that the check of 
Eqs.~(2.13) and (2.14) cannot be done so straightforwardly.

In this section, we will show that we can approximately check 
Eq.~(2.13) and (2.14) by using the values of $m_f$ and $V_{ij}$ 
at $\mu=m_Z$, without knowing the explicit values of $m_f$ and 
$V_{ij}$ at $\mu=\Lambda_X$, as far as the evolutions of
$m_f$ and $V_{ij}$ are not singular.

It is well known that in such a conventional model the evolution 
effects are approximately described as \cite{evol}
\begin{eqnarray}
&&\frac{m_u^0/m_t^0}{m_u/m_t} \simeq \frac{m_c^0/m_t^0}{m_c/m_t}
\simeq 1+ \varepsilon_u \ ,  \nonumber\\
&&\frac{m_d^0/m_b^0}{m_d/m_b} \simeq \frac{m_s^0/m_b^0}{m_s/m_b}
\simeq 1+ \varepsilon_d \ ,  \nonumber\\
&&\frac{|V_{ub}^0|}{|V_{ub}|} \simeq \frac{|V_{cb}^0|}{|V_{cb}|}
\simeq \frac{|V_{td}^0|}{|V_{td}|} \simeq \frac{|V_{ts}^0|}{|V_{ts}|}
\simeq 1+ \varepsilon_d \ ,  \nonumber\\
&&\frac{m_u^0/m_c^0}{m_u/m_c} \simeq \frac{m_d^0/m_s^0}{m_d/m_s}
\simeq \frac{|V_{us}^0|}{|V_{us}|} \simeq \frac{|V_{cd}^0|}{|V_{cd}|}
\simeq 1 \ , 
\end{eqnarray}
where $m_q^0$ and $V_{ij}^0$ ($m_q$ and $V_{ij}$) denote the  values 
at $\mu=\Lambda_X$ ($\mu=m_Z$). The relations (3.2) hold only for a 
model where the Yukawa coupling constant of top quark,
$y_t\equiv (Y_u)_{33}$, 
satisfies $y_t\gg (Y_d)_{ij}$ ($i,j=1,2,3$).
The relations (3.2) also hold 
even in a model which has an intermediate energy scale \(\Lambda_I\),
because, for example, when we denote $(m_u/m_t)_{\mu=\Lambda_X}/(m_u/
m_t)_{\mu=\Lambda_I}$ and $(m_u/m_t)_{\mu=\Lambda_I}/(m_u/m_t)_{\mu=
m_Z}$ as \(1+ \)\(\varepsilon_{u1}\) and \(1+ \)\(\varepsilon_{u2}\), 
respectively,
we can obtain $(m_u/m_t)_{\mu=\Lambda_X}/(m_u/m_t)_{\mu=m_Z} \simeq 
1+\varepsilon_u$ with $\varepsilon_u=\varepsilon_{u1}+\varepsilon_{u2}$.

By using the approximate relations (3.2) the diagonalized 
up-quark mass matrix $D_u^0$ at $\mu=\Lambda_X$ 
is presented as
\begin{eqnarray}
D_u^0 &= &m_t^0 \left(
\begin{array}{ccc}
m_u^0/m_t^0 & 0 & 0 \\
0 & m_c^0/m_t^0 & 0 \\
0 & 0 & 1 
\end{array} \right)
\nonumber \\
&\simeq & m_t^0 \left(
\begin{array}{ccc}
m_u/m_t & 0 & 0 \\
0 & m_c/m_t & 0 \\
0 & 0 & 1
\end{array} \right) \left(
\begin{array}{ccc}
1+\varepsilon_u & 0 & 0 \\
0 & 1+\varepsilon_u  & 0 \\
0 & 0 & 1
\end{array} \right) \nonumber \\
& = & \frac{m_t^0}{m_t} (1+ \varepsilon_u S) D_u ,
\end{eqnarray}
where
\begin{equation}
S=\left(
\begin{array}{ccc}
1 & 0 & 0 \\
0 & 1 & 0 \\
0 & 0 & 0 
\end{array} \right)\ .
\end{equation}
Similarly, the matrix $D_d^0$ is given by
\begin{equation}
D_d^0 \simeq   \frac{m_b^0}{m_b} (1+ \varepsilon_d S) D_d .
\end{equation}
The CKM matrix $V^0$ at $\mu=\Lambda_X$ is given by
\begin{eqnarray}
V^0 & \simeq & \left(
\begin{array}{ccc}
1 & V_{us} & V_{ub}(1+\varepsilon_d) \\
V_{cd} & 1 & V_{cb}(1+\varepsilon_d) \\
V_{td}(1+\varepsilon_d) & V_{ts} (1+\varepsilon_d) & 1 
\end{array} \right)
\nonumber \\
& \simeq & ({\bf 1}+\varepsilon_d S_3) V ({\bf 1}+\varepsilon_d S_3)
-2 \varepsilon_d S_3\ ,
\end{eqnarray}
where $S_3={\bf 1}-S$ and ${\bf 1}$ is a $3\times 3$ unit matrix.
By using the relations (3.4) - (3.6), we can obtain 
the approximate expression
\begin{equation}
V^0 D_d^0 V^{0T} \simeq \frac{m_b^0}{m_b}\left[ (1+\varepsilon_d) V D_d V^T
-\varepsilon_d m_b S_3 \right] ,
\end{equation}
where we have used the observed hierarchical relations among
the quark mass ratios and CKM matrix parameters.
Therefore, the matrix $V D_d V^T + \kappa D_u$ in 
Eqs.~(2.10)-(2.12) is given by
\begin{eqnarray}
K^0 & \equiv & V^0 D_d^0 V^{0T} + \kappa^0 D_u^0 
\nonumber \\
& \simeq & (1+\varepsilon_d)\frac{m_b^0}{m_b}\left(
V D_d V^T +\kappa D_u \right.
\nonumber \\
 & &  \left. -\varepsilon_d m_b S_3 +
\varepsilon_u \kappa D_u S \right)\ ,
\end{eqnarray}
where 
\begin{equation}
\kappa = \frac{m_t^0/m_t}{m_b^0/m_b} 
\frac{\kappa^0}{1+\varepsilon_d}\ .
\end{equation}
Since the solutions $\kappa$ are of the order of $10^{-2}$
as we show in the next section, 
we can neglect
the term $\kappa D_u S$ compared with $V D_d V^T$
(note that in order to neglect  the component $(D_u S)_{11}$
it is essential that the sign of $m_d/m_s$ is positive,
because $(V D_d V^T)_{11} \simeq m_d +V_{us}^2 m_s$ and
$V_{us}^2 \simeq |m_d/m_s|$). 
On the other hand, for such a small value of $\kappa$, the
term $m_b S_3$ cannot be neglected compared with the term
$\kappa D_u$.
However, for a small value of $\varepsilon_d$, we can find 
that the solutions $\kappa$ are substantially not affected 
by the term $\varepsilon_d m_b S_3$.
As a result, we obtain the approximate expression
\begin{equation}
K^0 \simeq (1+\varepsilon_d)\frac{m_b^0}{m_b}\left(
V D_d V^T +\kappa D_u \right) .
\end{equation}
Therefore, Eq.~(2.13) and (2.14) at $\mu=\Lambda_X$, i.e.,
\begin{equation}
\frac{[(m_e^0)^2+(m_\mu^0)^2+(m_\tau^0)^2]^3}{(m_e^0)^2
(m_\mu^0)^2(m_\tau^2)^2}=\frac{[{\rm Tr}(K^0 K^{0\dagger}])]^3}{
{\rm det}(K^0 K^{0\dagger})} \ ,
\end{equation}
\begin{eqnarray}
&& \frac{[(m_e^0)^2+(m_\mu^0)^2+(m_\tau^0)^2]^2}{2[(m_e^0)^2
(m_\mu^0)^2+(m_\mu^0)^2(m_\tau^0)^2+(m_\tau^0)^2(m_e^0)^2]} \\
\nonumber
&& =
\frac{[{\rm Tr}(K^0 K^{0\dagger})]^2}{[{\rm Tr}(K^0 K^{0\dagger})]^2-
{\rm Tr}(K^0K^{0\dagger})^2} \ ,
\end{eqnarray}
are approximately replaced by the relations at $\mu=m_Z$:
\begin{equation}
\frac{(m_e^2+m_\mu^2+m_\tau^2)^3}{m_e^2 m_\mu^2 m_\tau^2}
=
\frac{ [{\rm Tr}(KK^\dagger)]^3}{{\rm det}(KK^\dagger)}\ ,
\end{equation}
\begin{eqnarray}
&& \frac{(m_e^2+m_\mu^2+m_\tau^2)^2}{2(m_e^2 m_\mu^2 +m_\mu^2
m_\tau^2+m_\tau^2m_e^2)} \\
\nonumber
&& =\frac{[{\rm Tr}(KK^\dagger)]^2}{{\rm Tr}
[(KK^\dagger)]^2-{\rm Tr}(KK^\dagger)^2} \ ,
\end{eqnarray}
where
\begin{equation}
K=VD_dV^\dagger+\kappa D_u \ ,
\end{equation}
and $\kappa$ is given by Eq.~(3.9).
This means that when we find the solution $\kappa$ 
at $\mu=m_Z$, the solution at $\mu=\Lambda_X$ also exists,
no matter whether the model is  a SUSY one or a non-SUSY one.
Then, we can obtain the value $\kappa^0$ at $\mu=\Lambda_X$ 
from the relation (3.9) with the solution $\kappa$ at $\mu=m_Z$.

\section{Numerical study at $\mu=m_Z$}
As mentioned in the preceding section, 
if the solution \(\kappa\) exists at the energy scale $\mu=m_Z$,
the one at $\mu=\Lambda_X$ also exists.
Therefore, we investigate the relations (2.13) and (2.14) at $\mu=m_Z$.
Note that Eqs.(\ref{eq82511}) and (\ref{eq82512}) are realized by GUT scale  
because Eq.(\ref{eq90102}) is broken at \(\mu=m_Z\).
In the present section, tentatively, we assume that 
the Yukawa coupling constant $Y_{10}$ and $Y_{126}$ at \(\mu=m_Z\) keep 
their forms symmetrical, so that we can put the observed values 
$D_u$, $D_d$ and $V$ at $\mu=m_Z$ into the relations (\ref{eq82511}) 
and (\ref{eq82512}). 
For the fermion masses at \(\mu=m_Z\), we use the following values:
\cite{koide}
\begin{eqnarray}
&&m_t=181\pm13            \ {\rm GeV}, \hspace{0.9cm}
  m_b=3.00 \pm 0.11       \ {\rm GeV}, \nonumber\\
&&m_c=677^{+56}_{-61}     \ {\rm MeV}, \hspace{1.1cm}
  m_s=93.4^{+11.8}_{-13.0}\ {\rm MeV}, \nonumber\\
&&m_u=2.33^{+0.42}_{-0.45}\ {\rm MeV}, \hspace{0.7cm}
  m_d=4.69^{+0.60}_{-0.66}\ {\rm MeV}, \label{e060701} \\
&&m_{\tau}=1746.7 \pm 0.3 \ {\rm MeV}, \nonumber \\
&&m_{\mu}=102.75138 \pm 0.00033\ {\rm MeV}, \nonumber\\
&&m_e=0.48684727\pm0.00000014\ {\rm MeV}. \nonumber
\end{eqnarray}
The input values for the CKM matrix parameters have been taken as \cite{PRD}
\begin{eqnarray}
\theta_{12}&=&0.219-0.226,\hspace{0.5cm} \theta_{23}=0.037-0.043, \nonumber\\
\theta_{13}&=&0.002-0.005,
\end{eqnarray}
where
\end{multicols}
\widetext
\hspace{-0.5cm}
\rule{8.7cm}{0.1mm}\rule{0.1mm}{2mm}
\begin{equation}
V=
\left(\begin{array}{ccc}
c_{13}c_{12} & c_{13}s_{12} & s_{13}e^{-i\delta} \\
-c_{23}s_{12}-s_{23}c_{12}s_{13}e^{i\delta} & 
c_{23}c_{12}-s_{23}s_{12}s_{13}e^{i\delta} & 
s_{23}c_{13} \\
s_{23}s_{12}-c_{23}c_{12}s_{13}e^{i\delta} & 
-s_{23}c_{12}-c_{23}s_{12}s_{12}s_{13}e^{i\delta} & 
c_{23}c_{13}
\end{array}\right),
\end{equation}
\hspace{9.2cm}
\rule[-2mm]{0.1mm}{2mm}\rule{8.7cm}{0.1mm}
\begin{multicols}{2}
\narrowtext
\hspace{-1.5em}
with \(c_{ij}\equiv \cos \theta_{ij}\) and \(s_{ij}\equiv \sin \theta_{ij}\).
The calculation has been performed allowing all the combinations 
of the quark mass signatures. 
Here it should be noted that, since \(m_u\) is 
much smaller than \(m_c\) and \(m_t\), 
the difference of the sign of \(m_u\) 
scarcely makes a change of allowed regions.
In this calculation, 
we have selected \(\theta_{23}\) and \(\delta\) as input parameters 
and \(m_s\), \(c_d\) and \(\kappa\) as output parameters 
because the calculation is sensitive to these parameters.
We give the numerical results in Fig 1.
Here, except for \(m_s\), \(\theta_{23}\) and \(\delta\), 
we have adopted the center values of Eq.(\ref{e060701}) as input values.
Moving \(\theta_{23}\) at intervals of 0.0005 rad and fixing \(\delta=60{\rm ^\circ}\),
we search the solutions where \(\kappa_A\) and \(\kappa_B\) become coincident.
Our numerical analysis shows that
the solutions exist in the combinations of Table \ref{T-0}.
In a table \ref{tab91601}, 
we show the nearest solution of \(m_s\), \(\theta_{23}\) and \(\delta\) 
to the center values of Eq.(\ref{e060701}).

In the following we perform data fitting 
for the case of top line of Table \ref{tab91601}.
Eqs. (\ref{eq82501})-(\ref{eq82503}) can constrain only 
the absolute value of \(c_d\). 
The argument of the parameter \(c_d\) may be decided by taking neutrino sector 
into consideration in the future.
For the time being, 
we set \(c_d \equiv |c_d| e^{i \sigma } =  \, e^{0.107i}\) 
so that \(c_0\) becomes a real number:
\begin{eqnarray}
c_0 &=& \frac{1 - c_d}{c_u}=34.7, \\
c_1 &=& -\frac{3 + c_d}{c_u}=101.8- 10.8i.
\end{eqnarray}
In this case, the mass matrices in MeV are
\end{multicols}
\widetext
\hspace{-0.5cm}
\rule{8.7cm}{0.1mm}\rule{0.1mm}{2mm}
\begin{eqnarray}
M_0 &=& \frac{3V D_d V^T
          + c_d(\kappa D_u + V D_d V^T)}{4} 
     = \left(
        \begin{array}{ccc}
	-12.4 -  0.7 i & -23.0 -  1.8 i &    9.6 -  13.2 i \\
	-23.0 -  1.8 i & -91.5 -  3.9 i &  194.0 +  10.5 i \\
	  9.6 - 13.2 i & 194.0 + 10.5 i & 1874.9 - 180.0 i 
        \end{array}
        \right), \\
M_1 &=& \frac{V D_d V^T
         - c_d(\kappa D_u + V D_d V^T)}{4}
     = \left(
        \begin{array}{ccc}
	 4.19 + 0.69 i&   7.68 +  1.43 i&   -3.72 +   4.09 i\\
	 7.68 + 1.43 i&  24.14 +  3.88 i&  -65.05 -  10.48 i\\
	-3.72 + 4.09 i& -65.05 - 10.48 i& 1119.67 + 179.98 i
        \end{array}
        \right).
\end{eqnarray}
\hspace{9.2cm}
\rule[-2mm]{0.1mm}{2mm}\rule{8.7cm}{0.1mm}
\begin{multicols}{2}
\narrowtext
\hspace{-0.33cm}Here, using the condition 
\(\sqrt{|v_0^u|^2+|v_0^d|^2+|v_1^u|^2+|v_1^d|^2}\)\(=246\)GeV, 
we can get VEV's as
\begin{equation}
v_0^d= \frac{246\mbox{\,[GeV]}}{\sqrt{(|c_0|^2+1)+(|c_1|^2+1)|\rho|^2}}
\end{equation}
with \(\rho\equiv v_1^d/v_0^d\).
Then, the Yukawa couplings about {\bf 10} and {\bf 126} become 
\begin{equation}
Y_{10} = \frac{M_0}{v_0^d} , \quad
Y_{126} = \frac{M_1}{v_1^d}. \label{eq092701}
\end{equation}
We consider that the model should be calculable perturbativly.
We can see that every element of the Yukawa coupling constants (\ref{eq092701}) 
is smaller than one if we take a suitable value of \(|\rho|\).


\section{10 and 120}
In the SO(10) GUT scenario, we can also discuss the model 
with one {\bf 10} and one {\bf 120} by the same method.
The Yukawa couplings of {\bf 10} and {\bf 120} 
are symmetric and antisymmetric, respectively. 
If we consider a case that the Yukawa coupling constants of {\bf 10} 
are real and {\bf 120} pure imaginary, we can make them Hermitian, i.e., 
\(Y_{10}^\dagger = Y_{10}\) and \( Y_{120}^\dagger = Y_{120}\).
Therefore, by considering the real vacuum expectation values \(v_{10}\) and \(v_{120}\),
we can obtain the Hermitian mass matrices \(M_u\), \(M_d\) and \(M_e\):
\begin{eqnarray}
M_d&=&M_0+ M_2,\ \ M_e=M_0-3 M_2, \nonumber\\
M_u&=&c_0M_0+c_2M_2. \label{eq83101}
\end{eqnarray}
Then, we can diagonalize 
those by unitary matrices $U_u$, $U_d$ and $U_e$ as 
\begin{equation}
U_u^{\dagger}M_uU_u=D_u, \quad
U_d^{\dagger}M_dU_d=D_d, \quad
U_e^{\dagger}M_eU_e=D_e.
\end{equation}
Since the CKM matrix $V$ is given by 
\begin{equation}
V=U_u^{\dagger}U_d \ ,
\end{equation}
the relation (\ref{eq83101}) is re-written as follows:
\begin{equation}
(U_u^{\dagger}U_e)D_e(U_u^{\dagger}U_e)^{\dagger}
=c_{d}VD_{d}V^{\dagger} +c_uD_u.
\end{equation}
As stated previously, we have almost known 
the experimental values of $D_e$, $D_u$ and $VD_dV^{\dagger}$.  
Therefore, we obtain the independent three equations:
\begin{equation}
{\rm Tr}D_e=c_d[{\rm Tr}D_d+\kappa{\rm Tr} D_u],
\end{equation}
\begin{equation}
{\rm Tr}D_e^2=c_d^2[{\rm Tr}D_d^2+2\kappa{\rm Tr}(D_uVD_dV^{\dagger})
+\kappa^2{\rm Tr}D_u^2],
\end{equation}
\begin{equation}
{\rm det}D_e=c_d^3{\rm det}(VD_{d}V^{\dagger}+{\kappa}D_u),
\end{equation}
where $\kappa=c_u/c_d$.  For the parameter $\kappa$, we have two equations:
\begin{eqnarray}
&&{\frac{m_e^2+m_\mu^2+m_\tau^2}{(m_e+m_\mu+m_\tau)^2}} \nonumber\\
&&\qquad=
{\frac{{\rm Tr}D_d^2+2{\kappa}{\rm Tr}(D_{u}VD_{d}V^{\dagger})+
{\kappa}^2{\rm Tr}D_u^2}
{({\rm Tr}D_d+{\kappa}{\rm Tr}D_u)^2}}, \label{83102} \\
&&{\frac{m_e m_\mu m_\tau}{(m_e+m_\mu+m_\tau)^3}}
=
{\frac{{\rm det}(VD_{d}V^{\dagger}+{\kappa}D_u)}
{({\rm Tr}D_{d}+{\kappa}{\rm Tr}D_{u})^3}}. \label{83103}
\end{eqnarray}
Eqs. (\ref{83102}) and (\ref{83103}) are more simple than
Eqs. (\ref{eq82511}) and (\ref{eq82512}). 
\(c_d\) and \(\kappa\) are real
since we have assumed the \(M_u\), \(M_d\) and \(M_e\) to be Hermitian.
So the calculation is easier than the case for {\bf 10} and {\bf 126}. 
The numerical results are listed in Table \ref{T-1}-\ref{tab91605}.

\section{Summary and Discussion}
In conclusion, we have investigated whether an SO(10) model 
with two Higgs scalars can reproduce the observed mass spectra of 
the up- and down-quark sectors and charged lepton sector or not.
What is of great interest is to see whether we can find reasonable
values of the parameters $c_u$ and $c_d$ which satisfy the SO(10)
relation (2.5) or not.
For the case with one {\bf 10} and one {\bf 126} scalars, in a parameter
$\kappa= c_u/c_d$, we have obtained two equations (\ref{eq82511}) 
and (\ref{eq82512}) which hold at the unification scale
$\mu=\Lambda_X$ and which are described in terms of the observable
quantities (the fermion masses and CKM matrix parameters). 
We have sought for the solution of $\kappa$ approximately by using 
the observed fermion masses and CKM matrix parameters at 
$\mu= m_Z$ instead  of the observable quantities at $\mu=\Lambda_X$.
Although we have found no solution for real $\kappa$, we have
found four solutions for complex $\kappa$ which satisfy 
Eqs. (\ref{eq82511}) and (\ref{eq82512}) within the 
experimental errors.
Similarly, we have found four solutions for a model
with one {\bf 10} and one {\bf 120} scalars.
It should be worth while noting that the solutions in the latter 
model are real.
The latter model is very attractive because the origin of the CP
violation attributes only to the {\bf 120} scalar.
In the both models, we can make the magnitudes of all the Yukawa 
coupling constants smaller than one, so that the models are
safely calculable under the perturbation theory. 

By the way, note that the numerical results are very sensitive to 
the values of \(m_s\) and \(\theta_{23}\). 
For numerical fittings, it is favor that the strange quark mass \(m_s\)
is somewhat smaller than the center value $m_s=93.4$ MeV which is 
quoted in Ref.~\cite{koide}.

Also note that the relative sign of $m_d$ to $m_s$ in
each solution is positive, i.e, $m_d/m_s >0$ as seen in
Tables \ref{T-0} and \ref{T-1}.
It is well known that a model with a texture $(M_d)_{11}=0$ 
on the nearly diagonal basis of the up-quark mass matrix 
$M_u$ leads to the relation $|V_{us}|=\sqrt{{-m_d}/{m_s}}$
\cite{Vus}, where the relative sign is negative, i.e.,
$m_d/m_s <0$.
On the contrary, 
we can conclude that in the SO(10) model with two Higgs scalars, 
we cannot adopt a model with the texture $(M_d)_{11}=0$.

In the present paper, we have demonstrated that the 
unified description of the quark and charged lepton 
masses in the SO(10) model with two Higgs scalars is
possible.
However, we have not referred to the neutrino masses.
Concerning this problem, 
Brahmachari and Mohapatra have recently showed that one {\bf 10} 
and one {\bf 126} model is incompatible with large 
\(\nu_{\mu}\)-\(\nu_{\tau}\) mixing angle \cite{brahmachari}.
Since there are many possibilities  for neutrino mass generation 
mechanism, we are optimistic about this problem, too.
Investigating for a question whether an SO(10) model with two 
Higgs scalars can give a unified description of quark and lepton 
masses including neutrino masses and mixings or not is our 
next big task.

\clearpage
\narrowtext

\begin{table}
\begin{tabular}{c|ccc}
  num. & $(m_t, m_c, m_u)$ & $(m_b, m_s, m_d)$ & $(m_{\tau}, m_{\mu}, m_e)$
  \\ \hline
  (a)  & $(+ \ - \ +)$ & $(+\ - \ -)$ & $(+ \ \pm \ \pm)$ 
  \\
  (b)  & $(+ \ - \ -)$ & $(+\ - \ -)$ & $(+ \ \pm \ \pm)$ 
\end{tabular}
\caption{The combinations of the signs of $(m_t, m_c, m_u)$, 
$(m_b, m_s, m_d)$ and $(m_{\tau}, m_{\mu}, m_e)$. 
The notation \((m_t, m_c, m_u)\) \(=\) \((+\ -\ +)\) 
denotes $m_t>0$, $m_c<0$ and $m_u>0$. 
Eqs. (\ref{eq82511}) and (\ref{eq82512}) 
are not affected by the signs of charged leptons.}
\label{T-0}
\end{table}

\begin{table}[htbp]
\begin{tabular}{c|cc|ccc}
  & \multicolumn{2}{c|}{Input} & \multicolumn{3}{c}{Output} \\
  & \(|\theta_{23}|[{\rm rad}]\) &\(\delta[{\rm ^\circ}]\) 
  & \(m_s[{\rm MeV}]\) &  \(|c_d|\) & \(\kappa\) \\ \hline
(a) & 0.0420  & 60.
    & 76.3    & 3.15698 & \(-0.01928-0.00089 i\) \\
    & 0.0420  & 60.       
    & 76.3    & 3.03577 & \(-0.01937-0.00101 i\) \\
(b) & 0.0420  & 60.       
    & 76.3    & 3.13307 & \(-0.01929-0.00092 i\) \\
    & 0.0420  & 60.                           
    & 76.3    & 3.00558 & \(-0.01939-0.00105 i\) 
\end{tabular}
\caption{
Four sets of parameters giving good data fitting at \(\mu=m_Z\) 
for one {\bf 10} and one {\bf 126} Higgs scalars.
(a) and (b) correspond to the mass signatures in Table I, 
and the upper and lower lines do to the two intersections in Fig.1}
\label{tab91601}
\end{table}

\begin{table}
\begin{tabular}{c|ccc}
  num. & $(m_t, m_c, m_u)$ & $(m_b, m_s, m_d)$ & $(m_{\tau}, m_{\mu}, m_e)$
  \\ \hline
  (a-1)  & $(+ \ - \ +)$ & $(+\ - \ -)$ & $(+ \ + \ +)$ 
  \\
  (a-2)  & $(+ \ - \ +)$ & $(+\ - \ -)$ & $(+ \ + \ -)$ 
  \\
  (b-1)  & $(+ \ - \ -)$ & $(+\ - \ -)$ & $(+ \ + \ +)$ 
  \\
  (b-2)  & $(+ \ - \ -)$ & $(+\ - \ -)$ & $(+ \ + \ -)$ 
\end{tabular}
\caption{The combinations of the signs of \((m_t,\) \(m_c, m_u)\), 
$(m_b, m_s, m_d)$ and $(m_{\tau}, m_{\mu}, m_e)$ for 
one {\bf 10} and one {\bf 120} Higgs scalars.}
\label{T-1}
\end{table}

\begin{table}[htbp]
\begin{tabular}{c|cc|ccc}
  & \multicolumn{2}{c|}{Input} & \multicolumn{3}{c}{Output} \\
  & \(|\theta_{23}|[{\rm rad}]\) &\(\delta[{\rm ^\circ}]\) 
  & \(m_s[{\rm MeV}]\) &  \(c_d\) & \(\kappa\) \\ \hline
(a-1) & 0.0415 & 60. & 79.551    & 0.05905 & \(-\)0.01957\\
(a-2) & 0.0415 & 60. & 79.238    & 0.06124 & \(-\)0.01942 \\
(b-1) & 0.0415 & 60. & 79.673    & 0.05855 & \(-\)0.01960\\
(b-2) & 0.0415 & 60. & 79.316    & 0.06080 & \(-\)0.01945\\
\end{tabular}
\caption{
Four sets of parameters giving good data fitting at \(\mu=m_Z\) 
for one {\bf 10} and one {\bf 120} Higgs scalars.
(a-i) and (b-i) correspond to the mass signatures in Table III.}
\label{tab91605}
\end{table}

\begin{figure}[htbp]
\begin{center}
\includegraphics{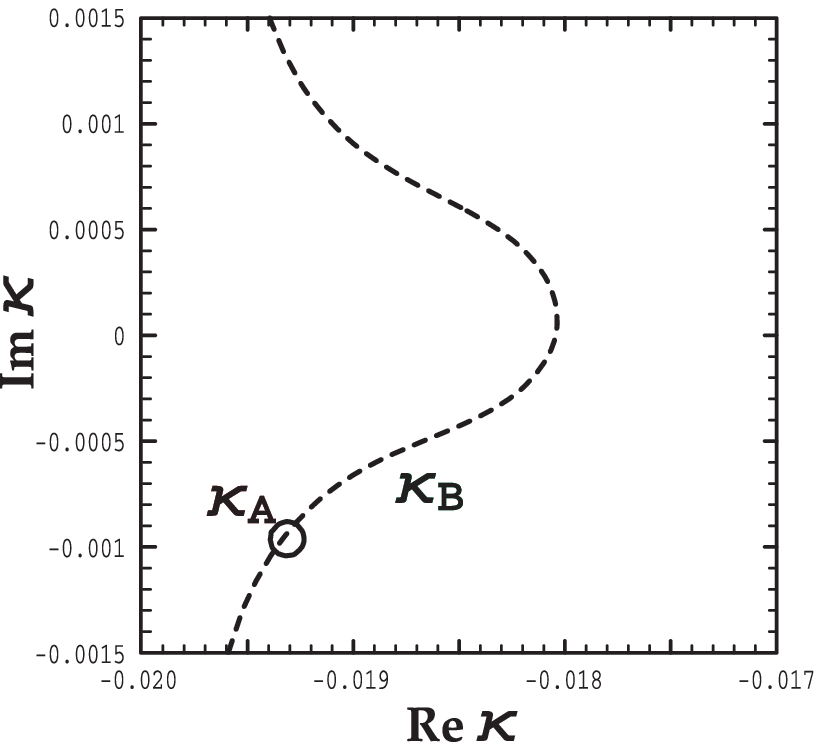}
\end{center}
\caption{The relations between Eqs.(2.13) and (2.14) 
on the complex plane of \(\kappa\).
The solid (dotted) line shows 
the solution of Eq.(2.13)  (Eq.(2.14)).}
\label{fig1}
\end{figure}


\end{multicols}

\end{document}